\documentclass[english]{article}
\usepackage{eucal}
\usepackage[utf8]{inputenc}
\usepackage[T1]{fontenc}
\usepackage{babel}
\usepackage{amsmath}
\usepackage{amsfonts}
\usepackage{csquotes}
\usepackage{amstext}
\usepackage{amssymb}
\usepackage{mathtools}
\usepackage{graphicx}
\usepackage{comment}
\usepackage{subcaption}
\usepackage{fancyhdr}
\usepackage{siunitx}
\usepackage{hyperref}
\hypersetup{
    colorlinks=true,
    linkcolor=blue,
    filecolor=magenta,      
    urlcolor=cyan,
}
\usepackage{graphicx}

\usepackage{biblatex}
\begin{document}

\title{Deep Neural Network Based Differential Equation Solver for HIV Enzyme Kinetics}

\begin{comment}
\author{Parvathy Jayan$^1$ \footnote{Equal contribution}
  \and Joseph Stember$^2$ $^*$\thanks{Corresponding author} 
  \and Hrithwik Shalu$^3$ }
\end{comment}

\author{ \textbf{Joseph Stember}$^1$
\thanks{Equal contribution} 
\and
\textbf{Parvathy Jayan}$^2$ $^*$
\and 
\textbf{Hrithwik Shalu}$^3$}

\maketitle

\thispagestyle{fancy}

\noindent
\textsuperscript{1}Memorial Sloan Kettering Cancer Center, New York, NY, US, 10065 
\\
\textsuperscript{2}Indian Institute of Science Education and Research, Tirupati, India, 517507
\\
\textsuperscript{3}Indian Institute of Technology, Madras, Chennai, India, 600036
\\

\noindent
\textsuperscript{1}joestember@gmail.com
\\
\textsuperscript{2}parvathyseena1@gmail.com
\\
\textsuperscript{3}lucasprimesaiyan@gmail.com 
\\

\begin{abstract}

\indent \textit{Purpose}  
We seek to use neural networks (NNs) to solve a well-known system of differential equations describing the balance between T cells and HIV viral burden. 

\indent \textit{Materials and Methods} 
In this paper we employ a 3-input parallel NN to approximate solutions for the system of first order ordinary differential equations describing the above biochemical relationship. 

\indent \textit{Results}
The numerical results obtained by the NN are very similar to a host of numerical approximations from the literature.

\indent \textit{Conclusion}
We have demonstrated use of NN integration of a well-known and medically important system of first order coupled ordinary differential equations. Our trial-and-error approach counteracts the system's inherent scale imbalance. However, it highlights the need to address scale imbalance more substantively in future work. Doing so will allow more automated solutions to larger systems of equations, which could describe increasingly complex and biologically interesting systems. 

\end{abstract}
\pagebreak

\section*{Introduction}

Differential equations underpin essentially all of science, engineering, and finance. They describe how the behavior of dynamical systems unfolds over time. Most differential equations cannot be solved symbolically, i.e., a closed form solution does not exist. Almost all systems that are of interest for studying practical applications must therefore be approximated numerically. Studying these systems centers on devising the appropriate numerical approximation scheme balancing the benefit of being able to obtain a solution with the deviation from a "true" answer that the approximation introduces.

With the recently surging popularity of deep learning and neural networks (NNs), new interest has arisen in solving differential equations with NNs. In the 1990s, Lagaris et al. proposed and implemented as proof-of-principle NN-based solutions to some illustrative ordinary and partial differential equations \cite{Lagaris_1998}. Dockhorn \cite{dockhorn2019discussion} applied the approach to solving the Poisson and Navier-Stokes Equations. Liu et al. \cite{liu2019neural} applied a NN approach to solving the Laplace Equation. In quantum mechanics, Sehanobish et al. \cite{sehanobish2021learning} used NNs to compute solutions for the potential energy function from Schrodinger's Equation. In classical mechanics, Mattheakis et al. \cite{mattheakis2020hamiltonian} solved Hamilton's Equations for positions and momenta, besting the fidelity of numerical solution phase space diagrams for both periodic and chaotic dynamical systems.

Chemical kinetics describes the rate of chemical conversions. It is an integral component in systems biology, in which large biochemical systems are studied quantitatively in order to understand biological behavior and disease processes, also forming an important tool in pharmacology. This is a rich field with meaningful applications in basic science and health care. As such, we seek to apply NNs to solve the differential equations that describe chemical kinetics. 

We endeavor specifically to do so for a particularly well known chemical kinetics system, a model for the interplay of human immunodeficiency virus (HIV) and immune system CD4+T cells, which the virus invades and attacks. This is the mechanism by which HIV degrades the human immune system. Particularly before the widespread use of highly active anti-retroviral therapy (HAART), this rendered victims susceptible to opportunistic infections that would normally be asymptomatic in the presence of a healthy immune system. 

\pagebreak

\section*{Methods}

\subsection*{Model and parameters}

The HIV kinetic model seeks to quantify the relative amounts / concentrations of the uninfected but susceptible T cells (not all T cells are susceptible to infection), infected T cells, and HIV virus particles over time. These are denoted by $T(t)$, $I(t)$, and $V(t)$, respectively. 

The system of ordinary first order differential equations connecting these variables can be formulated as \cite{atangana2014computational,perelson1999mathematical}: 

\begin{equation} \label{eq:diff_eqn}
\begin{aligned}
& \frac{dT}{dt} = p - \alpha T + rT\left( 1 - \frac{T + I}{T_{max}} \right) - kVT\\
& \frac{dI}{dt} =   kVT - \beta I   \\
& \frac{dV}{dt} =  N\beta I - \gamma V
\end{aligned}
\end{equation}

\noindent
subject to the initial conditions:

\begin{equation}
    T(0) = T_0,\qquad I(0) = I_0,\qquad V(0) = V_0
\end{equation}

Regarding the parameters in Equation \ref{eq:diff_eqn}:
\begin{itemize}
  \item $p$ is the rate of T cell production in the bone marrow and thymus
  \item $\alpha$ is the natural turnover rate of uninfected T cells
  \item $r$ is the rate of T cell mitosis, or division
  \item $T_{max}$ is the maximum concentration of T cells in the bloodstream
  \item $k$ is the rate constant for infection by the HIV virus
  \item $N$ is the number of infectious free viral particles (virions) produced per infected T cell \cite{perelson1993dynamics}
  \item $\beta$ is the natural turnover rate of infected T cells
  \item $\gamma$ is the natural turnover rate of virus particles
\end{itemize}

We seek to solve Equation \ref{eq:diff_eqn} using the experimentally known quantities \cite{atangana2014computational,ongun2011laplace,atangana2013solving,yuzbacsi2012numerical}:

\begin{equation}
\begin{aligned}
& T_0 = 0.1,\quad I_0 = 0.0,\quad V_0 = 0.1,\quad p = 0.1,\\ 
& \alpha = 0.02,\quad \beta=0.3,\quad \gamma = 2.4,\quad r = 3.0,\\ 
& k = 0.0027,\quad T_{max} = 1500,\quad N = 10
\end{aligned}
\end{equation}

We solve via NN loss minimization. In order to do this, we first subtract off the right hand sides of Equation \ref{eq:diff_eqn}:

\begin{equation} \label{eq:diff_eqn_set_eq_to_zero}
\begin{aligned}
& \frac{dT}{dt} - \left( p - \alpha T + rT\left( 1 - \frac{T + I}{T_{max}} \right) - kVT \right) = 0 \\
& \frac{dI}{dt} - ( kVT - \beta I ) = 0     \\
& \frac{dV}{dt} - ( N\beta I - \gamma V ) =  0
\end{aligned}
\end{equation}

\subsection*{Addressing scale imbalance}

As seen in Figure \ref{fig:function_plots} as well as earlier numerical results (Tables \ref{tab:T} \textendash \ref{tab:I}), in general, $T(t) \gg V(t) \gg I(t)$. More specifically, $T(t)$ values are on the order of $10^{-1}$\textendash  $10^0$, whereas $V(t)$ takes values on the order of $10^{-1}$ \textendash $10^{-2}$, with $I(t)$ much lower at around $10^{-5}$ \textendash $10^{-6}$. In general, the order of magnitude of a function being approximated by a NN is reflected in the network's loss.

Hence, a loss value that is significant enough to update weights for the NN approximating $I(t)$ would have a negligible effect upon that for $V(t)$, and the latter would fail to train / update, producing the vanish gradient problem. On the other hand, a loss significant enough to update $V(t)$ would have an outsized effect on the weights for $I(t)$, engendering the exploding gradient problem.

In order to ameliorate this problem, we multiply the $\frac{dI}{dt}$ and $\frac{dV}{dt}$ by scaling factors in order to bring all quantities toward the same order of magnitude. Based on some trial and error, as well as the relative typical size scale of the values as mentioned above, the modified set of differential equations that we solve with the NN from Figure \ref{fig:CNN_architecture} is given by:
\begin{equation} \label{eq:diff_eqns_scaled}
\begin{aligned}
    & \frac{dT}{dt} - \left( p - \alpha T + rT\left( 1 - \frac{T + I}{T_{max}} \right) - kVT \right) = 0 \\
& 10,000 \times \left( \frac{dI}{dt} - ( kVT - \beta I ) \right) = 0     \\
& 10 \times \left( \frac{dV}{dt} - ( N\beta I - \gamma V ) \right) =  0 
\end{aligned}
\end{equation}
Two major drawbacks to this approach are:

\begin{enumerate}
    \item It requires a process of trial and error that can not be automated. Furthermore, this would soon become unfeasible for even slightly larger systems of differential equations.
    \item Equation \ref{eq:diff_eqn} represents a \textit{coupled} system of differential equations. For example, knowledge of any of $I(t)$ and $V(t)$ is required to calculate $T(t)$. This interdependence means that we can not change (via multiplication) any lines of Equation \ref{eq:diff_eqn} without altering all of the functional values. Hence, although it works in this example, the approach of Equation \ref{eq:diff_eqns_scaled} is not in general a reliable one for bringing functions to the same general order of magnitude.
\end{enumerate}

\subsection*{Neural network}

Having made the above adjustments, our loss will be the right hand side of Equation \ref{eq:diff_eqns_scaled}. We can enforce the equality to arbitrary accuracy by minimizing the loss to within a particular threshold. For the time derivative terms, we substitute the finite difference numerical approximations:

\begin{equation} \label{eq:finite_diff}
\begin{aligned}
    \frac{dT}{dt} \approx \frac{T(t_i + \Delta t) - T(t_i)}{\Delta t} \\
    \frac{dI}{dt} \approx \frac{I(t_i + \Delta t) - I(t_i)}{\Delta t} \\
    \frac{dV}{dt} \approx \frac{V(t_i + \Delta t) - V(t_i)}{\Delta t},
\end{aligned}
\end{equation}

\begin{figure}[h!]
\centering
\includegraphics[width=12cm,height=12cm]{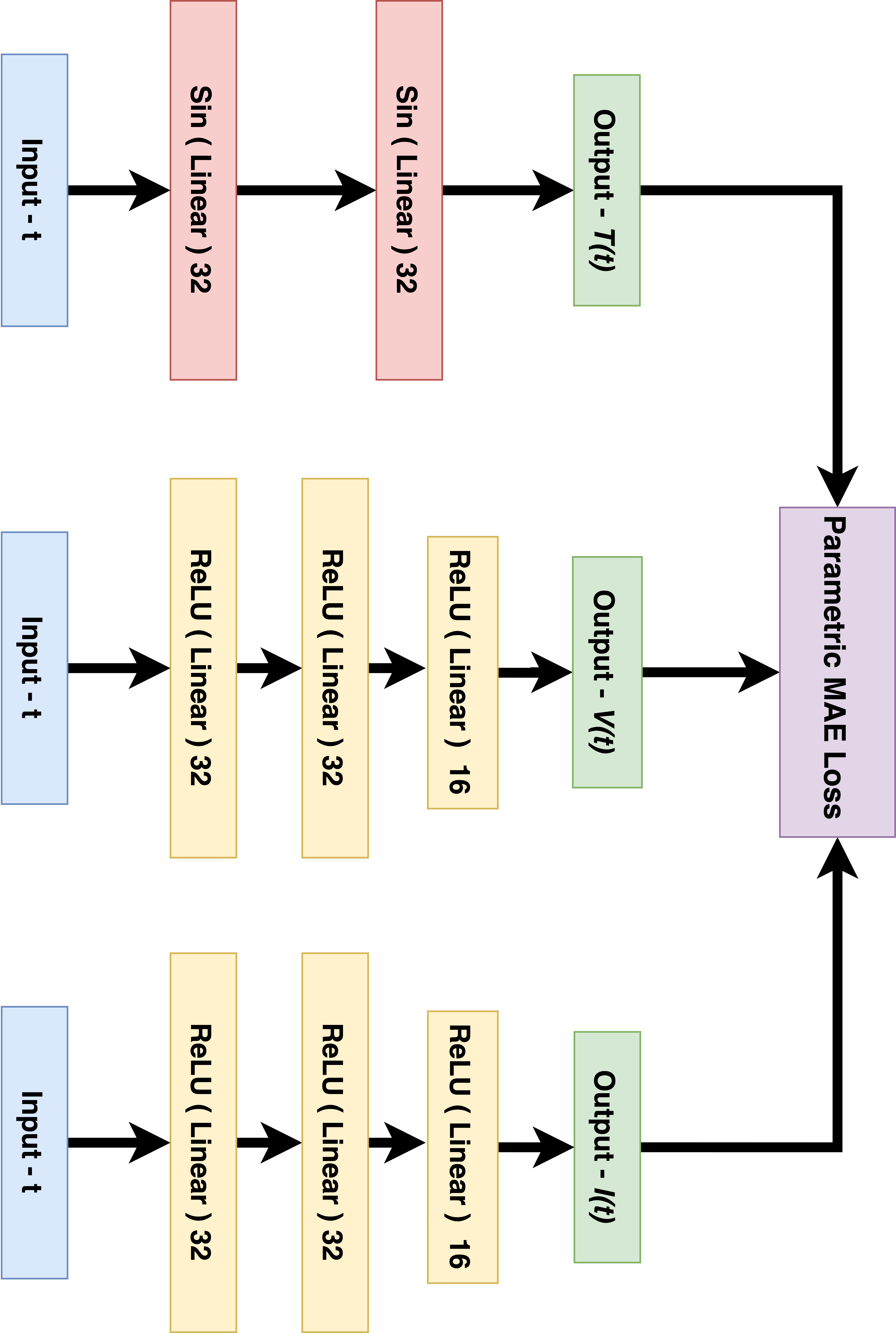}
\caption{ Deep neural network architecture. }
\label{fig:CNN_architecture}
\end{figure}

where $t_i$ is the time value with which we estimate the slope and $\Delta t$ is a small time increment. We note that Equation \ref{eq:finite_diff} becomes strict equalities as $\Delta t \rightarrow 0$. Hence, the smaller the value we choose for $\Delta t$, the better the approximation. 

The one input into our neural network is time $t$. The domain on which the function is defined is given by the span of $t$ values on which the network is trained. The architecture of the relatively simple NN we employ here is shown in Figure \ref{fig:CNN_architecture}. The input $t$ is passed separately through a parallel set of network branches. We use the three parallel branches / multi-input structure because we are approximating three different functions. 

for the $T(t)$ branch, this consists of two fully connected layers of 32 nodes, each with sine activation. The sine function normalizes to between -1 and 1, ensuring that network weights do not explode or vanish. It is used for the $T(t)$ branch because $T(t)$ is on a higher scale of magnitude compared to $I(t)$ and $V(t)$. 

Since $I(t)$ and $V(t)$ take much lower values, their NN branches use ReLu activation. ReLu could, given sizable input values, produce large values in the intermediate layers or loss. However, this is not of concern given typically low values of $I(t)$ and $V(t)$. 

As they tend to be of lower magnitude, $I(t)$ and $V(t)$ require more approximation power, and thus have an additional fully connected layer of 16 nodes before producing the single node outputs (representing the scaler function values $I(t)$ and $V(t)$.) 

Hence, the three branches together produce a 3-node output. Each of the output nodes, $T(t)$, $I(t)$, and $V(t)$ are compared to the numerical values of Equation \ref{eq:diff_eqns_scaled} using the approximations from Equation \ref{eq:finite_diff}. We use mean absolute error for our loss function. 

We employ the following parameters / hyperparameters in training our NN:

\begin{itemize}
  \item Adam optimizer
  \item Learning rate: $1 \times 10^{-3}$
  \item Training time: 3,000 epochs
  \item Training set size: 256 time values
  \item Testing set size: 128 time values
\end{itemize}

The input time values $t_s$ that the network samples span from $t_{\text{min}}=0$ to $t_{\text{max}}=1$. Outside of interval $\left[ 0, 1 \right]$, the NN will not learn how to fit the functions. We were able to sample within this interval with distinct training and validation sets as follows:
\begin{itemize}
    \item Training set: selecting $t$ values from a uniform random distribution.
    \item Testing set: selecting $t$ values from a uniformly spaced distribution; all $t$s generated in the same epoch will form a grid where each $t$ is equally spaced.
\end{itemize}

All calculations were executed in Google Colab Pro, which runs in the Python language. We made use of the Pytorch module for NN design and training, also using the neurodiffeq library, which is specifically designed for solving differential equations with NNs. 

\section*{Results}

\begin{figure}[h!]
\centering
\includegraphics[width=12cm,height=12cm]{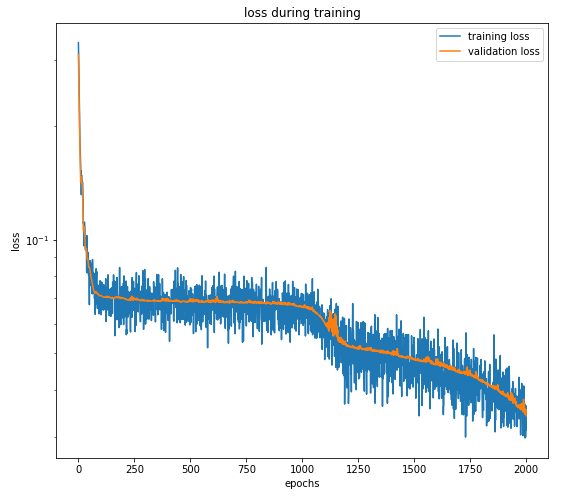}
\caption{ Loss of the neural network as a function of training time (epochs). }
\label{fig:loss}
\end{figure}

The loss during training is shown in Figure \ref{fig:loss}. Effective training is manifested by monotonically decreasing loss of both the training and validation sets.

\begin{figure}[h!]
\centering
\advance\leftskip-3cm
\includegraphics[width=18cm,height=17cm]{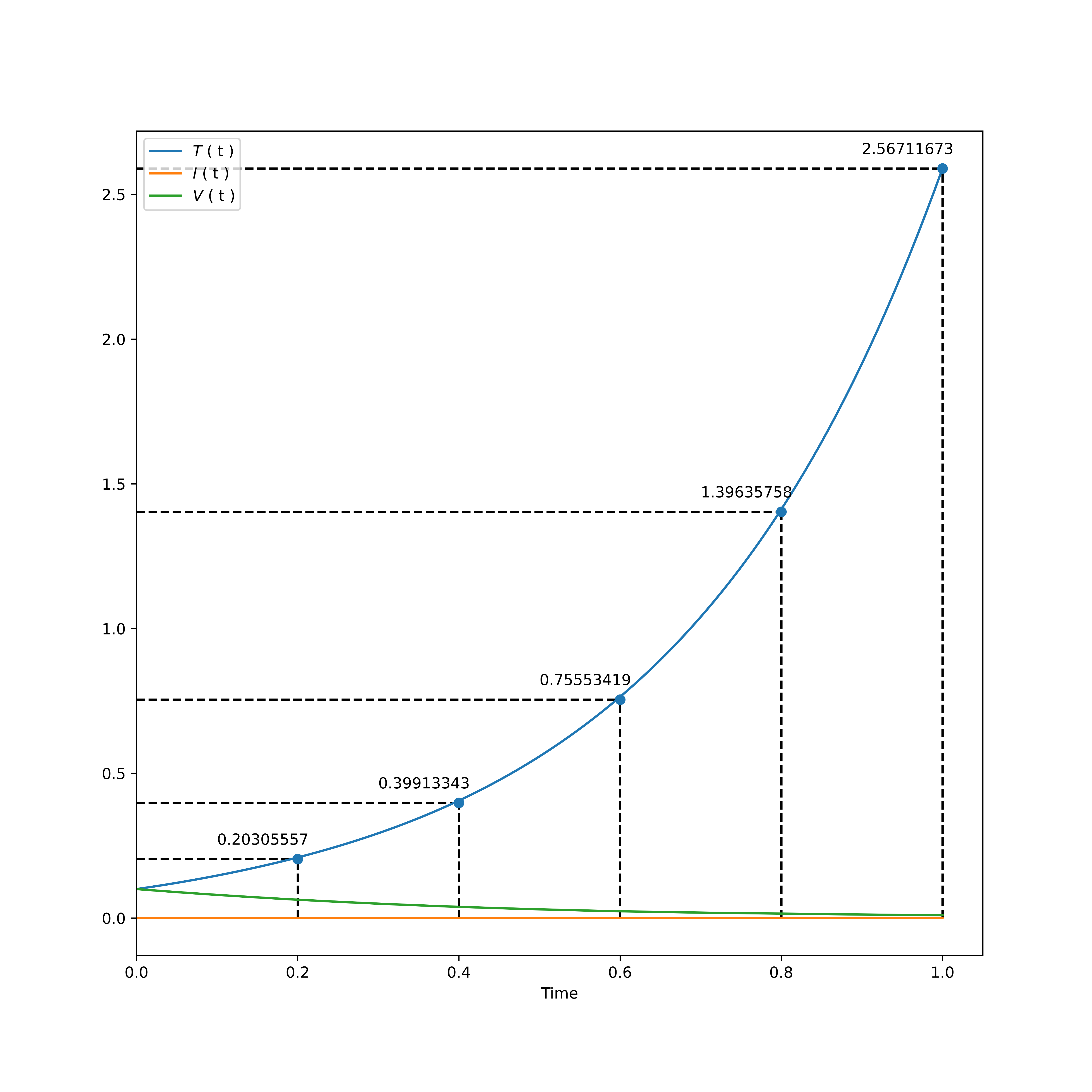}
\caption{Plots of the solved concentrations over time.}
\label{fig:function_plots}
\end{figure}

The computed values for $T(t)$, $I(t)$, and $V(t)$ are shown in Figure \ref{fig:function_plots}. The expected trend of $T(t)$ increasing over time and being of a higher order of magnitude compared to $I(t)$ and $V(t)$ is apparent. The overall trend matches well with previously published results \cite{atangana2013solving}. 

By identifying the interpolated values of $T(t)$ at particular time values, we can compare with numerical approximation methods previously used to approximate the solution to Equation \ref{eq:diff_eqn}. The NN values are tabulated in the second-to-left-most column in Table \ref{tab:T}. They are juxtaposed with the corresponding values from various leading numerical approximation methods, and we can see that the values are very close. It should be noted that, despite the apparent increase in T cells implied by Figure \ref{fig:function_plots}, this represents a sub-population of T cells on a relatively short time scale. This is in fact bounded by overall T cell depletion at a longer time scale due to viral destruction \cite{perelson1993dynamics}, as would be expected, and noting in fact that one definition of AIDS is total CD4 count below 200 $\frac{\text{cells}}{mm^3}$.

\section*{Conclusions}

If we were to use the numerical approximations as gold standards, the above would be a demonstration of high accuracy. We know from prior work that NN-based integration is in fact more accurate than numerical approximate solutions and is often indistinguishable from the results of analytical integration in cases where a closed form solution is possible \cite{mattheakis2020hamiltonian}. 

An important limitation of the current approach is the aforementioned issue of scale imbalance. Although we achieved good results for this system with a trial and error approach that involved multiplying two of the equations by scaling factors, this is not guaranteed to work on other systems. In fact, the lack of a systematic method to achieve scale balance would be expected to hinder integration for larger, more complex systems that would be of interest in, for instance, systems biology. Future work will focus on new techniques to overcome this challenge.

\begin{table}[h!]
\centering
\advance\leftskip-2cm
\begin{tabular}{llllllll}
\hline \\
$t$   & \begin{tabular}[c]{@{}l@{}}Proposed\\ Method\end{tabular} & HDM\cite{atangana2014computational}        & LADM-Padé\cite{ongun2011laplace}  & Runge-Kutta & MVIM\cite{atangana2013solving}        & VIM\cite{atangana2013solving}         & BCM\cite{yuzbacsi2012numerical}          \\ \\ \hline \\
0   & 0.1                                                       & 0.1        & 0.1        & 0.1         & 0.1        & 0.1        & 0.1        \\ \\ 
0.2 & 0.20305557                                                & 0.20880727 & 0.20880727 & 0.20880808  & 0.20880808 & 0.20880732 & 0.20386165 \\ \\
0.4 & 0.39913343                                                & 0.40610526 & 0.40610526 & 0.40624053  & 0.40624079 & 0.40613465 & 0.38033093 \\ \\
0.6 & 0.75553419                                                & 0.76114677 & 0.76114677 & 0.76442388  & 0.76442872 & 0.76245303 & 0.69546237 \\ \\
0.8 & 1.39635758                                                & 1.37731985 & 1.37731985 & 1.41404683  & 1.41409417 & 1.39788058 & 1.27596244 \\ \\
1   & 2.56711673                                                & 2.32916976 & 2.32916976 & 2.59159480  & 0.20880808 & 2.50674666 & 2.38322774 \\ \\ \hline
\end{tabular}
\caption{Numerical comparison of $T(t)$}
\label{tab:T}
\end{table}

\begin{table}[h!]
\centering
\advance\leftskip-2cm
\begin{tabular}{llllllll}
\hline \\
$t$   & \begin{tabular}[c]{@{}l@{}}Proposed\\ Method\end{tabular} & HDM\cite{atangana2014computational}        & LADM-Padé\cite{ongun2011laplace}  & Runge-Kutta & MVIM\cite{atangana2013solving}       & VIM\cite{atangana2013solving}         & BCM\cite{yuzbacsi2012numerical}         \\ \\ \hline \\
0   & 0.1                                                       & 0.1        & 0.1        & 0.1         & 0.1        & 0.1        & 0.1        \\ \\
0.2 & 0.06334023                                                & 0.06187996 & 0.06187996 & 0.06187984  & 0.06187990 & 0.06187995 & 0.06187991 \\ \\
0.4 & 0.03875031                                                & 0.03831324 & 0.03831324 & 0.03829488  & 0.03829595 & 0.03830820 & 0.03829493 \\ \\
0.6 & 0.02408340                                                & 0.02439174 & 0.02439174 & 0.02370455  & 0.02371029 & 0.02392029 & 0.02370431 \\ \\
0.8 & 0.01513445                                                & 0.00996721 & 0.00996721 & 0.01468036  & 0.01470041 & 0.01621704 & 0.01467956 \\ \\
1   & 0.00962319                                                & 0.00330507 & 0.00330507 & 0.00910084  & 0.00915723 & 0.01608418 & 0.02370431 \\ \\ \hline
\end{tabular}
\caption{Numerical comparison of $V(t)$}
\label{tab:V}
\end{table}

\begin{table}[h!]
\centering
\advance\leftskip-3.3cm
\begin{tabular}{llllllll}
\hline \\
$t$   & \begin{tabular}[c]{@{}l@{}}Proposed\\ Method\end{tabular} & HDM\cite{atangana2014computational}            & LADM-Padé\cite{ongun2011laplace}      & Runge-Kutta    & MVIM\cite{atangana2013solving}            & VIM\cite{atangana2013solving}             & BCM\cite{yuzbacsi2012numerical}            \\ \\ \hline \\
0   & 0                                                         & 0              & 0              & 0              & 0.1 · 10\textsuperscript{-13}    & 0              & 0              \\ \\
0.2 & 8.18247 · 10\textsuperscript{-6}                                            & 6.03270 · 10\textsuperscript{-6} & 6.03270 · 10\textsuperscript{-6} & 6.03270 · 10\textsuperscript{-6} & 6.03270 · 10\textsuperscript{-6} & 6.03263 · 10\textsuperscript{-6} & 6.24787 · 10\textsuperscript{-6} \\ \\
0.4 & 1.52065 · 10\textsuperscript{-5}                                            & 1.31591 · 10\textsuperscript{-5} & 1.31591 · 10\textsuperscript{-5} & 1.31583 · 10\textsuperscript{-5} & 1.31583 · 10\textsuperscript{-5} & 1.31487 · 10\textsuperscript{-5} & 1.29355 · 10\textsuperscript{-5} \\ \\
0.6 & 2.34348 · 10\textsuperscript{-5}                                            & 2.12683 · 10\textsuperscript{-5} & 2.12683 · 10\textsuperscript{-5} & 2.12237 · 10\textsuperscript{-5} & 2.12233 · 10\textsuperscript{-5} & 2.10141 · 10\textsuperscript{-5} & 2.03526 · 10\textsuperscript{-5} \\ \\
0.8 & 3.09622 · 10\textsuperscript{-5}                                            & 3.00691 · 10\textsuperscript{-5} & 3.00691 · 10\textsuperscript{-5} & 3.01774 · 10\textsuperscript{-5} & 3.01745 · 10\textsuperscript{-5} & 2.79513 · 10\textsuperscript{-5} & 2.83730 · 10\textsuperscript{-5} \\ \\
1   & 3.85396 · 10\textsuperscript{-5}                                            & 3.98736 · 10\textsuperscript{-5} & 3.98736 · 10\textsuperscript{-5} & 4.00378 · 10\textsuperscript{-5} & 4.00254 · 10\textsuperscript{-5} & 2.43156 · 10\textsuperscript{-5} & 3.69084 · 10\textsuperscript{-5} \\ \\ \hline
\end{tabular}
\caption{Numerical comparison of $I(t)$}
\label{tab:I}
\end{table}

\pagebreak

\printbibliography

\end{document}